\documentstyle[amsfonts,aps]{revtex}

\begin{document}
\author{Mario Castagnino}
\address{CONICET-Instituto de Astronom\'\i a y F\'\i sica del Espacio\\
Casilla de Correos 67, Sucursal 28, 1428, Buenos Aires, Argentina}
\author{Olimpia Lombardi}
\address{CONICET-Universidad Aut\'onoma de Madrid\\
Ctra. Colmenar Km 15, 28049, Madrid, Espa\~{n}a}
\title{A global and non-entropic arrow of time: the double role of the
energy-momentum tensor}
\maketitle

\begin{abstract}
In this paper we adopt a global and non-entropic approach to the problem of
the arrow of time, according to which the arrow of time is an intrinsic
geometrical property of spacetime. Our main aim is to show the double role
played by the energy-momentum tensor in the context of our approach. On the
one hand, the energy-momentum tensor is the intermediate step that permits
to turn the geometrical time-asymmetry of the universe into a local arrow of
time manifested as a time-asymmetric energy flow. On the other hand, the
energy-momentum tensor supplies the basis for deducing the time-asymmetry of
quantum field theory, posed as an axiom in this theory.
\end{abstract}

\section{Introduction}

Since the nineteenth century, the problem of the arrow of time has been one
of the most controversial questions in the foundations of physics and one of
the main concerns of many physicists. In the last years we were very
interested on this problem and have devoted several papers to the
cosmological origin of the arrow of time (\cite{Casta Varios}). This
research has reached its culmination in papers \cite{Olimpia} and \cite
{Foundations}, where we have presented a comprehensive formulation of our
view on the subject based on a global and non-entropic approach, according
to which the arrow of time is an intrinsic geometrical property of spacetime.

The main aim of this paper is to show the double role played by the
energy-momentum tensor in the context of our approach. On the one hand, the
energy-momentum tensor is the intermediate step that allows us to turn the
geometrical time-asymmetry of the universe into a local arrow of time
manifested as a time-asymmetric energy flow. On the other hand, the
energy-momentum tensor supplies the basis for deducing the time-asymmetry of
quantum field theory posed as an axiom in this theory.

When the problem of the arrow of time is addressed, the main obstacle to be
faced is conceptual confusion: the lack of consensus is primarily due to the
fact that different concepts are identified and different problems are
subsumed under the same label. Thus, it is not possible to seek an
acceptable solution if the terms used are not adequately defined in physical
or in mathematical precise terms. For this reason we are forced to devote
the initial sections of the paper to clarify several concepts and to
disentangle the problems involved in the discussion in order to reach our
goal avoiding misunderstandings and misinterpretations. On this basis, the
paper is organized as follows. In Section 2 the problems of irreversibility
and of the arrow of time are precisely stated and distinguished. In Section
3 the global and non-entropic character of our approach is justified by
contrast with other traditional approaches. In Section 4 the necessary
conditions for defining a global and non-entropic arrow of time are
explained and applied to the case of FLRW\ models. In Section 5 it is shown
the role played by the energy-momentum tensor in translating the global
geometrical arrow into the local level. Finally, in Section 6 it is shown
how the energy-momentum tensor can be used to justify the time-asymmetry
postulate of quantum field theory on global grounds.

\section{Disentangling problems: irreversibility and arrow of time}

Traditionally, the problem of irreversibility and the problem of the arrow
of time have been identified, as if irreversibility were the clue for
understanding the origin and the nature of the arrow of time. In the present
section we will show that, when the concepts involved in the debate are
precisely clarified, both problems become evidently different. In
particular, we will see that the question of irreversibility presupposes the
previous answer of the problem of the arrow of time.

\subsection{The problem of irreversibility}

In general, the concepts of irreversibility and of time-reversal invariance
are invoked in the discussions about the arrow of time, but usually with no
elucidation{\bf \ }of their precise meanings; this results in confusions
that contaminate many interesting considerations. For this reason, we will
start from providing some necessary definitions.

\begin{quote}
{\bf Definition 1: }A dynamical equation is {\it time-reversal invariant} if
it is invariant under the transformation $t\rightarrow -t$; as a result, for
each solution $f(t)$, $f(-t)$ is also a solution.

{\bf Definition 2:} A solution of a dynamical equation is {\it reversible}
if it corresponds to a closed curve in phase space.
\end{quote}

It is quite clear that both concepts are different to the extent that they
apply to different mathematical entities: time-reversal invariance is a
property of dynamical equations (laws) and, {\it a fortiori}, of the set of
its solutions (evolutions); reversibility is a property of a single solution
of a dynamical equation. Furthermore, they are not even correlated. In fact,
time-reversal invariant equations can have irreversible solutions\footnote{%
For instance, the equation of motion of the pendulum with Hamiltonian:
\par
\[
H=\frac{1}{2}\,p_{\theta }^{2}+\frac{K^{2}}{2}\,\cos \theta 
\]
is {\it time-reversal invariant}, namely, it is invariant under the
transformation $\theta \rightarrow \theta ,$ $p_{\theta }\rightarrow
-p_{\theta }$; however, whereas the trajectories within the separatrices are
reversible since they are closed curves, the trajectories above (below) the
separatrices are {\it irreversible} since, in the infinite time-limit, $%
\theta \rightarrow -\infty $ ($\theta \rightarrow +\infty )$.}.

When both concepts are elucidated in this way, {\it the problem of
irreversibility} can be clearly stated: {\it how to explain irreversible
evolutions in terms of time-reversal invariant laws}. But once it is
recognized that irreversibility and time-reversal invariance apply to
different mathematical entities, it is easy to find a conceptual answer to
the problem of irreversibility: nothing prevents a time-reversal invariant
equation from having irreversible solutions.

Of course, even though the conceptual answer is simple, a great deal of
theoretical work is needed for obtaining irreversible evolutions from an
underlying time-reversal invariant dynamics. This was the problem faced by
the founding fathers of statistical mechanics when they sought to describe
the irreversible evolutions of thermodynamics by means of the time-reversal
invariant laws of classical mechanics. At present, many efforts are directed
to explain the irreversible behavior of quantum systems in terms of the
time-reversal invariant quantum theory (see, for instance, \cite{Irrev}).
Here we only mean that, in order to face the problem of irreversibility, the
question about the arrow of time does not need to be addressed: the
distinction between the two directions of time is {\it usually presupposed}
when the irreversible evolutions are conceived as processes going from
non-equilibrium to equilibrium or from preparation to measurement towards
the future.

\subsection{The problem of the arrow of time}

In our everyday life we perceive an asymmetry between past and future and
experience the time order of the world as ''directed''. The problem of the
arrow of time arises when we seek a physical correlate of this intuitive
asymmetry. The main difficulty to be encountered for solving this problem is
our anthropocentric perspective, which prevents us from shaking off our
temporally asymmetric assumptions. In fact, traditional discussions around
the problem of the arrow of time are usually subsumed under the label ''the
problem of the direction of time'', as if we could find an exclusively
physical criterion for singling out {\it the} direction of time, identified
with what we call ''the future''. But there is nothing in physical laws that
distinguishes, in a non-arbitrary way, between past and future as we
conceive them in everyday life. It might be objected that physics implicitly
assumes this distinction with the use of temporally asymmetric expressions,
like ''future light cone'', ''initial conditions'', ''increasing time'', and
so on. However this is not the case, and the reason relies on the
distinction between ''conventional'' and ''substantial''.

\begin{quote}
{\bf Definition 3: }Two objects are {\it formally identical} when there is a
permutation that interchanges the objects but does not change the properties
of the system to which they belong.
\end{quote}

In physics it is usual to work with formally identical objects: the two
semicones of a light cone, the two spin senses, etc. Now we can define two
notions that will be central in the further discussion:

\begin{quote}
{\bf Definition 4:} We will say that we establish a {\it conventional}
difference when we call two formally identical objects with two different
names, e.g{\it .}, when we assign different signs to the two spin senses.

{\bf Definition 5:} We will say that the difference between two objects is 
{\it substantial} when we assign different names to two objects which are
not formally identical (see \cite{Penrose}, \cite{Sachs}). In this case,
even though the names are conventional, the difference is substantial.
\end{quote}

E.g., the difference between the two poles of the theoretical model of a
magnet is conventional since both poles are formally identical; on the
contrary, the difference between the two poles of the Earth is substantial
because in the north pole there is an ocean and in the south pole there is a
continent (and the difference between ocean and continent remains
substantial even if we conventionally change the names of the poles).

Once this point is accepted, it turns to be clear that physics uses the
labels ''past'' and ''future'' in a conventional way. Therefore, the problem
cannot yet be posed in terms of singling out the future direction of time:
the problem of the arrow of time becomes the problem of finding a {\it %
substantial difference} between the two temporal directions. But if this is
our central question, we cannot project our independent intuitions about
past and future for solving it without begging the question. When we want to
address the problem of the arrow of time from a perspective purged of our
temporal intuitions, we must avoid the conclusions derived from subtly
presupposing time-asymmetric notions. As Huw Price \cite{Price} claims, it
is necessary to stand at a point outside of time, and thence to regard
reality in atemporal terms: this is ''{\it the view from nowhen}''. This
atemporal standpoint prevents us from using temporally asymmetric
expressions in a non-conventional way: the assumption about the difference
between past and future is not yet legitimate in the context of the problem
of the arrow of time.

But then, what does ''the arrow of time'' mean when we accept this
constraint? Of course, the traditional expression coined by Eddington has
only a metaphorical sense: its meaning must be understood by analogy. We
recognize the difference between the head and the tail of an arrow on the
basis of its geometrical properties; therefore, we can substantially
distinguish between both directions, head-to-tail and tail-to-head,
independently of our particular perspective. Analogously, we will conceive 
{\it the problem of the arrow of time} in terms of {\it the possibility of
establishing a substantial distinction between the two directions of time on
the basis of exclusively physical arguments}.

\section{A global and non-entropic approach}

\subsection{Why global? The traditional local approach}

The traditional local approach owes its origin to the attempts to reduce
thermodynamics to statistical mechanics: in this context, the usual answer
to the problem of the arrow of time consists in defining the future as the
direction of time in which entropy increases. However, already in 1912 Paul
and Tatiana Ehrenfest \cite{Ehrenfest} noted that, if the entropy of a
closed system increases towards the future, such increase is matched by a
similar one in the past of the system. Gibbs' answer to the Ehrenfests'
challenge was based on the assumption that probabilities are determined{\it %
\ from prior events to subsequent events}. But this answer clearly violates
the ''nowhen'' standpoint since probabilities are blind to temporal
direction. Therefore, any appeal to the distinction between prior and
subsequent events commits a {\it petitio principii} by presupposing the
arrow of time from the very beginning.

The point can also be posed in different terms. Let us assume that we have
solved the irreversibility problem; so we have the description of all the
irreversible evolutions, say, decaying processes, of the universe. However,
since we have not yet established a substantial difference between both
directions of time, we have no way to decide towards which temporal
direction {\it each} decay proceeds. Of course, we would obtain the arrow of
time if we could coordinate the processes in such a way that all of them
parallelly decay towards the same temporal direction. But this is precisely
what local physics cannot offer: only by means of global considerations all
the decaying processes can be coordinated. This means that the global arrow
of time plays the role of the background scenario where we can meaningfully
speak of the temporal direction of irreversible processes, and this scenario
cannot be built up by means of local theories that only describe phenomena
confined in small regions of spacetime.

\subsection{Why non-entropic? The traditional entropic approach}

When, in the late nineteenth century, Boltzmann developed the probabilistic
version of his theory in response to the objections raised by Loschmidt and
Zermelo, he had to face a new challenge: how to explain the highly
improbable current state of our world. In order to answer this question,
Boltzmann offered the first cosmological approach to the problem: ''{\it The
universe, or at least a big part of it around us, considered as a mechanical
system, began in a very improbable state and it is now also in a very
improbable state. Then if we take a smaller system of bodies, and we isolate
it instantaneously from the rest of the world, in principle this system will
be in an improbable state and, during the period of isolation, it will
evolve towards more probable states}'' \cite{Boltzmann}. Since this seminal
work, many authors have related the temporal direction past-to-future to the
gradient of the entropy function of the universe. For instance, Feynman
asserts: {\bf ''}{\it For some reason, the universe at one time had a very
low entropy for its energy content, and since then entropy has increased. So
that is the way towards future. That is the origin of all\ irreversibility}%
{\bf ''} \cite{Feynman}. In a similar sense, Davies claims that {\bf ''}{\it %
There exists an arrow of time only because the universe originates in a
less-than-maximum entropy state}{\bf ''} \cite{Davies}. Even if these
authors admit the need of global arguments for solving the problem of the
arrow of time, they coincide in considering that it must be addressed in
terms of entropy.

The global entropic approach rests on two assumptions: that it is possible
to define entropy for a complete cross-section of the universe, and that
there is an only time for the universe as a whole. However, both assumptions
involve difficulties. In the first place, the definition of entropy in
cosmology is still a very controversial issue: there is not a consensus
regarding how to define a global entropy for the universe. In fact, it is
usual to work only with the entropy associated with matter and radiation
because there is not yet a clear idea about how to define the entropy due to
the gravitational field. In the second place, when general relativity comes
into play, time cannot be conceived as a background parameter which, as in
pre-relativistic physics, is used to mark the evolution of the system.
Therefore, the problem of the arrow of time cannot legitimately be posed,
from the beginning, in terms of the entropy gradient of the universe
computed on a background parameter of evolution.

Nevertheless, these points are not the main difficulty: there is a
conceptual argument for abandoning the traditional entropic approach. As it
is well known, a given value of entropy is compatible with many
configurations of a system: entropy is a phenomenological property. The
question is whether there is a more fundamental property of the universe
which allows us to distinguish between both temporal directions. On the
other hand, if the arrow of time reflects a substantial difference between
both directions of time, it is reasonable to consider it as an intrinsic
property of time, or better, of spacetime, and not as a secondary feature
depending on a phenomenological property. For these reasons we will follow
Earman's ''{\it Time Direction Heresy}'' \cite{Earman}, according to which
the arrow of time is an intrinsic, geometrical property of spacetime
embodied in $g_{\mu \nu }(x)$, which does not need to be reduced to
non-temporal features. In other words, the geometrical approach to the
problem of the arrow of time has conceptual priority over the entropic
approach, since the geometrical properties of the universe are more basic
than its thermodynamic properties.

\section{Conditions for a global and non-entropic arrow of time}

\subsection{Time-orientability}

In a Minkowski spacetime, it is always possible to define the class of all
the future light semicones and the class of all the past light semicones
(where the labels ''future'' and ''past'' are conventional). In general
relativity the metric can always be approximated, in small regions of
spacetime, to the Minkowski form. However, on the large scale, we do not
expect the manifold to be flat because gravity can no longer be neglected.
Many different topologies are consistent with Einstein's field equations. In
particular, the possibility arises of spacetime being curved along the
spatial dimension in such a way that, e.g.{\bf , }the spacelike sections of
the universe become the three-dimensional analogous of a Moebius band; in
this case it is said that the spacetime is non time-orientable.

\begin{quote}
{\bf Definition 6:} A spacetime is {\it time-orientable} if there exists a
continuous non-vanishing vector field $\gamma ^{\mu }(x)$ on it which is
everywhere non-spacelike.
\end{quote}

By means of this field, the set of all semicones of the manifold can be
split into two equivalence classes, $C_{+}$ and $C_{-}$: the semicones of $%
C_{+}$ contain the vectors of the field and the semicones of $C_{-}$ do not
contain them. On the other hand, in a non time-orientable spacetime it is
possible to transform a future pointing timelike vector into a past pointing
timelike vector by means of a continuous transport that always keeps
non-vanishing timelike vectors timelike; therefore, the distinction between
future semicones and past semicones cannot be univocally definable on a
global level. This means that the time-orientability of spacetime is a
precondition for defining a global arrow of time, since if spacetime is not
time-orientable, it is not possible to distinguish between the two temporal
directions for the universe as a whole.

Nevertheless, not all accept this conclusion. For instance, Matthews \cite
{Matthews} claims that a spacetime may have a regional but not a global
arrow of time if the arrow is defined by means of local considerations.
However, even from this local approach (which we have rejected in the
previous section), time-orientability cannot be avoided. Let us suppose that
there were a local non time-reversal invariant law $L$, which defines
regional arrows of time that disagree when compared by means of continuous
timelike transport. The trajectory of the transport will pass through a
frontier point between both regions: in a region around this point the arrow
of time will be not univocally defined, and this amounts to a breakdown of
the validity of $L$ in such a point. But this fact contradicts the
methodological {\it principle of universality}, unquestioningly accepted in
contemporary cosmology, according to which the laws of physics are valid in
all points of the spacetime. The strategy to escape this conclusion would
consist in refusing to assign any meaning to the timelike continuous
transport. This strategy would only be acceptable if the two regions with
different arrows were physically isolated: this amounts to the
disconnectedness of the spacetime. But this fact would contradict another
methodological principle of cosmology, that is, the {\it principle of
uniqueness}, according to which there is only one universe and completely
disconnected spacetimes are not allowed\footnote{%
Although there are quantum cosmologies exhibiting disconnected space-times,
such models only play an explanatory role since they are not testably in
principle. Anyway, even if disconnected space-times were allowed, each
connected region could be considered as a universe by itself, where timelike
continuous transport must be valid. This fact is relevant since we are
interested in explaining the time direction of our own connected universe.
For a different opinion, see \cite{Schul} and our criticism in \cite{Olimpia}%
.}. These arguments show that the possibility of arrows of time pointing to
opposite directions in different regions of the spacetime is not an
alternative seriously considered in contemporary cosmology.

\subsection{Cosmic time}

As it is well known, general relativity replaces the older conception of
space-through-time by the concept of spacetime, where time becomes a
dimension of a four-dimensional manifold. But when the time measured by a
physical clock is considered, each particle of the universe has its own {\it %
proper time}, that is, the time registered by a clock carried by the
particle. Since the curved spacetime of general relativity can be considered
locally flat, it is possible to synchronize the clocks fixed to particles
whose parallel trajectories are confined in a small region of spacetime.
But, in general, the synchronization of the clocks fixed to all the
particles of the universe is not possible. Only in certain particular cases
all the clocks can be coordinated by means of a cosmic time, which has the
features necessary to play the role of the temporal parameter in the
evolution of the universe.

The issue can also be posed in geometrical terms. A spacetime may be such
that it is not possible to partition the set of all events into equivalent
classes such that: (i) each one of them is a spacelike hypersurface, and
(ii) the hypersurfaces can be ordered in time. There is a hierarchy of
conditions which, applied to a time-orientable spacetime, avoid
''anomalous'' temporal features (see \cite{Hawking-Ellis}). The strongest
condition is the existence of a global time.

\begin{quote}
{\bf Definition 7}: A {\it global time function} on the Riemannian manifold $%
M$ is a function $t:M\rightarrow {\Bbb R}$ whose gradient is everywhere
timelike.
\end{quote}

The value of the global time function increases along every future directed
non-spacelike curve. The existence of such a function guarantees that the
spacetime is globally splittable into hypersurfaces of simultaneity which
define a foliation of the spacetime (see \cite{Schutz}).

Nevertheless, the fact that the spacetime admits a global time function does
not yet permit to define this notion of simultaneity in an univocal manner
and with physical meaning. In order to avoid ambiguities in the notion of
simultaneity, we must choose a particular foliation. The foliation $\tau $
such that there exists a continuous set of worldline curves which are
orthogonal to all the hypersurfaces $\tau =const.$ is the proper choice,
because orthogonality recovers the notion of simultaneity of special
relativity for small regions (tangent hyperplanes) of the hypersurfaces $%
\tau =const.$ (for the necessary conditions, see \cite{Misner}). However,
even if this condition selects a particular foliation, it permits that the
proper time interval between two hypersurfaces of simultaneity be relative
to the particular worldline considered for computing it. If we want to avoid
this situation, we must impose an additional constraint: the proper time
interval between two hypersurfaces $\tau =\tau _{1}$ and $\tau =\tau _{2}$
must be the same when measured on any orthogonal worldline curve of the
continuous set mentioned above. In this case, the metric results: 
\begin{equation}
ds^{2}=dt^{2}-h_{ij}\,dx^{i}\,dx^{j}  \label{5.1}
\end{equation}

\begin{quote}
{\bf Definition 8}: When the metric of the spacetime can be expressed as $%
ds^{2}=dt^{2}-h_{ij}\,dx^{i}\,dx^{j}$, $t$ is the {\it cosmic time} and $%
h_{ij}=h_{ij}(t,x^{1},x^{2},x^{3})$ is the three-dimensional metric of each
hypersurface of simultaneity.
\end{quote}

Of course, the existence of cosmic time imposes a significant topological
and metric limitation on the spacetime. But with no cosmic time, there is
not a single time which can be considered as the parameter of the evolution
of the universe and, therefore, it is nonsensical to speak of the two
directions of time for the universe as a whole. Therefore, the possibility
of defining a cosmic time is a precondition for meaningfully speaking of a
global arrow of time. This fact supplies an additional argument against the
entropic approach, which takes for granted the possibility of defining the
entropy function of the universe. But this amounts to the assumption that:
(i) the spacetime can be partitioned in spacelike hypersurfaces on which the
entropy of the universe can be defined, and (ii) the spacetime possesses a
cosmic time on which the entropy gradient can be computed. When the
possibility of spacetimes with no cosmic time is recognized, it is difficult
to deny the conceptual priority of the geometrical structure of spacetime
over entropic features in the context of our problem.

\subsection{Time-asymmetry}

It is quite clear that time-orientability is merely a necessary condition
for defining the global arrow of time, but it does not provide a physical,
substantial criterion for distinguishing between the two directions of time.
As we will see, such a distinction requires the time-asymmetry of the
universe.

It is usually accepted that the obstacle to the definition of the arrow of
time lies in the fact that the fundamental laws of physics are time-reversal
invariant\footnote{%
The exception is the law that rules weak interactions. We will return on
this point in the last section.}. Nevertheless, this common position can be
objected when the concept of time-symmetry is clearly elucidated and
compared with the concept of time-reversal invariance: whereas time-reversal
invariance is a property of dynamical equations (laws), time-symmetry is a
property of a single solution (evolution) of an dynamical equation.

\begin{quote}
{\bf Definition 9}: A solution $f(t)$ of a dynamical equation is
time-symmetric if there is a time $t_S$ such that $f(t+t_S)=f(t-t_S)$.
\end{quote}

It is quite clear that the time-reversal invariance of an equation does not
imply the time-symmetry of its solutions: a time-reversal invariant law may
be such that all or most of the possible evolutions relative to it are
individually time-asymmetric. Huw Price \cite{Price} illustrates this point
with the familiar analogy of a factory which produces equal numbers of
left-handed and right-handed corkscrews: the production as a whole is
completely unbiased, but each individual corkscrew is asymmetric.

Although these considerations are not applicable to the field equations as
originally stated, the existence of a cosmic time allows to present the
issue in familiar terms: under this condition, Einstein's field equations
are time-reversal invariant in the sense that if the $%
h_{ij}(t,x^{1},x^{2},x^{3})$ of eq.(1) is a solution, $%
h_{ij}(-t,x^{1},x^{2},x^{3})$ is also a solution. But the time-reversal
invariance of these equations does not prevent us from describing a
time-asymmetric universe whose spacetime is asymmetric regarding its
geometrical properties along the cosmic time. This idea can also be
formulated in terms of the concept of time-isotropy.

\begin{quote}
{\bf Definition 10}: A time-orientable spacetime $(M,g)$ (where $M$ is a
four-dimensional Riemannian manifold and $g$ is a Lorentzian metric for $M$)
is {\it time-isotropic} if there is a diffeomorphism $d$ of $M$ onto itself
which reverses the temporal orientations but preserves the metric $g$.
\end{quote}

However, when we want to express the time-symmetry of a spacetime having a
cosmic time, it is necessary to strengthen the definition.

\begin{quote}
{\bf Definition 11}: A time-orientable spacetime which admits a cosmic time $%
t$ is {\it time-symmetric} with respect to some spacelike hypersurface $%
t=t_S $, where $t_S$ is a constant, if it is time-isotropic and the
diffeomorphism $d$ leaves fixed the hypersurface $t=t_S$.
\end{quote}

Intuitively this means that, from the hypersurface $t=t_{S}$, the spacetime 
{\it looks the same} in both temporal directions. Therefore, if a
time-orientable spacetime having a cosmic time is time-asymmetric, we will
not find a spacelike hypersurface $t=t_{S}$ which splits the spacetime into
two ''halves'', one the temporal mirror image of the other with respect to
their intrinsic geometrical properties.

But, how this time-asymmetry allows us to choose a time-orientation of the
spacetime? As we have seen, in a time-orientable spacetime a continuous
non-vanishing non-spacelike vector field $\gamma ^{\mu }(x)$ can be defined
all over the manifold. Up to this point, the universe is {\it time-orientable%
} but not yet {\it time-oriented}, because the distinction between $\gamma
^{\mu }(x)$ and $-\gamma ^{\mu }(x)$ is just conventional. Now
time-asymmetry comes into play: in a time-orientable time-asymmetric
spacetime, {\it any} time $t_{A}$ splits the manifold into two sections that
are different to each other: the section $t>t_{A}$ is {\it substantially}
different than the section $t<t_{A}$. We can choose any point $x_{0}$ with $%
t=t_{A}$ and conventionally consider that a certain non-spacelike vector{\bf %
\ }$-\gamma ^{\mu }(x_{0})$ points towards $t<t_{A}$ and that $\gamma ^{\mu
}(x_{0})$ points towards $t>t_{A}$ or vice versa: in any case we have
established a substantial difference between $\gamma ^{\mu }(x_{0})$ and $%
-\gamma ^{\mu }(x_{0})$. We can conventionally call the direction of $\gamma
^{\mu }(x_{0})$ ''future'' and the direction of $-\gamma ^{\mu }(x_{0})$
''past'' or vice versa, but in any case past is substantially different than
future. Now we can extend this difference to the whole continuous fields $%
\gamma ^{\mu }(x)$ and $-\gamma ^{\mu }(x)${\bf \ }obtained by making the
global continuations of $\gamma ^{\mu }(x_{0})$ and $-\gamma ^{\mu }(x_{0})$%
\ allowed by the definition of time-orientability: in this way, the
time-orientation of the spacetime has been established. Since the field $%
\gamma ^{\mu }(x)$ is defined all over the manifold, it can be used {\it %
locally }at each point $x$ to define the future and the past semicones: for
instance, if we have called the direction of $\gamma ^{\mu }(x)$ ''future'', 
$C_{+}(x)$ contains $\gamma ^{\mu }(x)$ and $C_{-}(x)$ contains $-\gamma
^{\mu }(x)$.

\subsection{Application to FLRW models}

In the previous subsections we have considered the general conditions
necessary for the existence of a global arrow of time. But, what can we say
about our universe? Cosmology offers a simple answer by means of the
cosmological principle and the assumption of expansion, on the basis of
which the metric of the spacetime can be represented in the simple
Robertson-Walker form: 
\begin{equation}
ds^2=dt^2-a^2(t)\left( \frac{dr^2}{1-kr^2}+r^2d\theta ^2+r^2\sin ^2\theta
d\phi ^2\right)  \label{FLRW}
\end{equation}
With the Robertson-Walker metric, the Einstein's field equations can be
solved: the corresponding solutions describe the isotropic and homogeneous
Friedmann-Lem\^aitre-Robertson-Walker models (FLRW for short), which are the
standard models of present day cosmology.

FLRW models provide a simple answer to the question about the first
condition for the existence of the arrow of time since they are
time-orientable. Moreover, astronomical observations provide empirical
evidence that makes implausible the non time-orientability of our spacetime.
In particular, there is no astronomical observation of temporally inverted
behavior in some (eventually very distant) region of the universe. In fact:

1.- The evolutions of several generations of supernovae always follow the
same pattern (let's say, from birth to death), and there is no trace of a
time-inverted pattern in the visible universe. At the decoupling time
(400.000 years after the Big-Bang) the universe was essentially composed of
light elements like H and He (with traces of isotopes of these elements and
of Li and Be) and was virtually free of ions. This matter condensed in
stars, where heavy elements were formed. The explosion of these stars as
supernovae scattered the heavy elements producing clouds which, in turn,
condensed giving rise to a second generation of stars. Analogously, a third
generation arose, and so on. We have an indirect evidence of the first
generation (known as population III), appeared 200 millions years after the
Big-Bang, by the observed re-ionization of the universe which can be
explained only by the presence of stars at that period \cite{WMAP}. Then,
two new generations (populations II and I) followed. This is the history as
manifest in all the universe, and this is a relevant fact in the context of
our problem since the supernovae are the markers or standard candles used to
measure the more distant galaxies ($z=1$ and even more) \cite{SUIe}.

2.- If we consider that the peak in the quasars formation rate took place
when the universe was 3.000 millions years old ($z\sim 2$), and that this
rate has always decreased since then, we see that the evolution of the
universe is time-asymmetric, at lest up to the corresponding distances.

3.- Finally, the process of decoupling between matter an radiation occurred
400.000 years after the Big-Bang ($z=1000$) and never happened again. This
fact proves the time-asymmetry of the visible universe (since 400.000 years
after the Big-Bang up to the present).

With regard to the existence of a cosmic time, since FLRW models are
spatially homogeneous and isotropic on the large scale, it is possible to
find a family of spacelike hypersurfaces which can be labeled by the proper
time of the worldlines that orthogonally thread through them: these labels
define the cosmic time, which is represented by the variable $t$, and the
scale factor $a$ is a scalar only function of $t$.

With regard to time-symmetry, in FLRW models the time-symmetry of spacetime
may manifest itself in two different ways according to whether the universe
has singular points in one or in both temporal extremities\footnote{%
This depends on the values of the factor $k$ and of the cosmological
constant $\Lambda $.}. Big Bang-Big Chill universes are manifestly
time-asymmetric: since the scale factor $a(t)$ increases with the cosmic
time $t$, there is no hypersurface $t=t_{S}$ from which the spacetime looks
the same in both temporal directions. In Big Bang-Big Crunch universes, on
the contrary, $a(t)$ has a maximum value: therefore, the spacetime might be
time-symmetric about the time of maximum expansion. These are the cases of
some FLRW models with dust and radiation. In more general cases ({\it e.g.}
inflationary models) different fields can represent the matter-energy of the
universe. Many interesting results have been obtained, for instance, by
modeling matter-energy as a set of scalar fields $\phi _{k}(t)$: homogeneity
is retained and the time-reversal invariance of the field equations is given
by the fact that, if $\left[ a(t),\phi _{k}(t)\right] $ is a solution, $%
\left[ a(-t),\phi _{k}(-t)\right] $ is also a solution. In these cases, if
there is a time $t_{ME}$ of maximum expansion, the scale factor $a(t)$ may
be such that $a(t_{ME}+t)\neq a(t_{ME}-t)$ (see, for instance, the models in 
\cite{Cast-Giac-Lara}). This means that a Big Bang-Big Crunch universe may
be a time-asymmetric object with respect to the metric of the spacetime, and
this asymmetry, essentially grounded on geometrical considerations, allows
us to distinguish{\bf \ }between the two directions of the cosmic time,
independently of entropic considerations.

As Savitt \cite{Savitt} correctly points out, there are two different
questions involved in the problem of the arrow of time:

\begin{itemize}
\item  The ''how possible'' question: how is it possible to formulate a
time-asymmetric model by means of time-reversal laws?

\item  The ''how probable'' question: what reason is there to suppose that
time-asymmetry is probable?
\end{itemize}

In the previous subsection we have shown that the time-reversal invariance
of laws is not an obstacle to the construction of time-asymmetric models.
However, the second question remains. The answer to this question can be
given by a simple argument that proves the vanishing probability of perfect
time-symmetry (see \cite{Olimpia}): such an argument demonstrates that
time-symmetric solutions of the universe equations have measure zero in the
corresponding phase space. This means that geometrical time-asymmetry is a
very specific feature which requires an overwhelmingly improbable
fine-tuning of all the state variables of the universe.

\section{From geometry to energy flow: the first role of the energy-momentum
tensor}

\subsection{The arrow of time as energy flow}

Up to this point, the arrow of time was defined in terms of the substantial
difference between the vector fields $\gamma ^{\mu }(x)$ and $-\gamma ^{\mu
}(x)$, grounded on the time-asymmetry of the spacetime. But $\gamma ^{\mu
}(x)$ was characterized merely as the vector field that must exist for the
time-orientability of spacetime. The question now is whether the arrow of
time can be defined in a physical way, that is, by means of a mathematical
object that can be interpreted not only geometrically but also in terms of
the more familiar magnitudes of physics.

As it is well known, the energy-momentum tensor $T_{\mu \nu }\ $represents
the density and the flow of energy and momentum at each point of the
spacetime. Then, it would be desirable to define the vector field $\gamma
^{\mu }(x)$ in terms of $T_{\mu \nu }$ in order to endow it with a physical
meaning. Although this task cannot be accomplished in a completely general
case, it is possible to define the arrow of time in terms of $T_{\mu \nu }$
when the energy-momentum tensor satisfies the {\it dominant energy condition}
everywhere (see \cite{Hawking-Ellis}, \cite{Visser}).

\begin{quote}
{\bf Definition 12}: The energy-momentum tensor satisfies the {\it dominant
energy condition} if, in any orthonormal basis, the energy component
dominates the other components of $T_{\alpha \beta }$: 
\[
T^{00}\geq \left| T^{\alpha \beta }\right| \qquad for\,\,each\,\,\alpha
,\beta 
\]
\end{quote}

This means that to any observer the local matter-energy density appears
non-negative and the energy flow is non-spacelike. The dominant energy
condition does not impose a very strong constraint, since it holds for
almost all known forms of matter\footnote{%
There are, of course, strange cosmological ''objects'' whose existence would
lead to universes where the dominant energy condition is not satisfied in
certain regions of spacetime. For instance, in wormhole spacetimes, the
dominant energy condition is violated in the vicinity of the wormhole throat
since the wormhole is threaded by negative ''exotic'' matter (see \cite
{Visser}). Nevertheless, it is plausible to suppose that universes
containing such kind of objects will surely not satisfy the stronger
conditions necessary for defining the arrow of time, that is,
time-orientability and existence of cosmic time.}.

Let us consider a continuous orthonormal basis field $\left\{ V_{(\alpha
)}^{\mu }(x)\right\} $ (a tetrad or {\it vierbein}) consisting of four
unitary vectors $(V_{(0)}^{\mu }(x),V_{(1)}^{\mu }(x),V_{(2)}^{\mu
}(x),V_{(3)}^{\mu }(x))$. In this basis, $g_{\mu \nu }V_{(\alpha )}^{\mu
}V_{(\beta )}^{\nu }=$ $\eta _{\alpha \beta }^{{}}$ are the coordinates of
the local Minkowski metric tensor, and $T_{\mu \nu }^{{}}V_{(\alpha )}^{\mu
}V_{(\beta )}^{\nu }=$ $T_{\alpha \beta }$ are the coordinates of the
energy-momentum tensor. Then, $T^{0\alpha }V_{(\alpha )}^{\mu }$ can be
conceived as a vector representing the energy flow, whose coordinates in
that basis are the $T^{0\alpha }$. Now, if $T^{00}\geq \left| T^{\alpha
\beta }\right| $, then $T^{00}\geq \left| T^{0\alpha }\right| $. In turn, $%
T^{00}\geq \left| T^{0\alpha }\right| $ implies that $T^{0\alpha }V_{(\alpha
)}^{\mu }$ is non-spacelike. On the other hand, if the manifold and the
basis field are continuous, $g_{\mu \nu }$ is continuously defined over it
and, provided that the derivatives of $g_{\mu \nu }$ are also continuous, $%
T^{\mu \nu }$ ($T^{\alpha \beta }$) and, then, $T^{0\mu }$ ($T^{0\beta }$)
are also continuously defined all over the manifold. Therefore, it seems
that, if the density of matter-energy is non-zero everywhere, we have found 
{\bf a} physical correlate of the continuous non-vanishing non-spacelike
vector field $\gamma ^{\mu }(x)=T^{0\alpha }(x)V_{(\alpha )}^{\mu }(x)$. The
trouble with this conclusion is that $T^{0\alpha }V_{(\alpha )}^{\mu }$ is
not strictly a vector, since it is not transformed as a vector by the
Lorentz transformations. Strictly speaking, at each point $x$ of the
spacetime $T^{0\alpha }(x)V_{(\alpha )}^{\mu }(x)$ is a tetra-magnitude
which represents the energy flow only in the basis $\left\{ V_{(\alpha
)}^{\mu }(x)\right\} $; thus, it seems that it cannot directly play the role
of $\gamma ^{\mu }(x)$ as initially desired.

Nevertheless, the fact that the energy flow cannot be represented by a
vector is not an obstacle to define the arrow of time in terms of such a
flow. The dominant energy condition poses a {\it covariant} condition: if
the energy flow is non-spacelike in a reference frame, it is non-spacelike
in all reference frames. This means that, no matter which orthonormal basis $%
\left\{ V_{(\alpha )}^{\mu }(x)\right\} $ is chosen, the energy flow in that
basis, represented by $T^{0\alpha }(x)V_{(\alpha )}^{\mu }(x)$, can be used
to define the arrow of time. The usual convention in physics consists in
calling the temporal direction of the positive energy flow ''future''. In
this case, at any point $x$ of the spacetime $T^{0\alpha }(x)V_{(\alpha
)}^{\mu }(x)$ belongs to the future light semicone $C_{+}(x)$: the energy
flows towards the future for any observer. But the relevant point is that
this sentence acquire a non-conventional meaning only when the substantial
difference between past and future has been previously established by the
time-asymmetry of the spacetime.

\subsection{From the global arrow to the local arrow}

As we have seen, the future light semicone $C_{+}(x)$ at each point $x$ of
the spacetime is defined by the positive energy flow $T^{0\alpha
}(x)V_{(\alpha )}^{\mu }(x)$ at this point. But, is $T^{0\alpha }$ really
the energy flow as conceived by local physics? Let us remember that $T_{\mu
\nu }$ satisfy the ''conservation'' equation:

\[
\nabla _{\mu }\,T^{\mu \nu }=0 
\]
However, this is not a true conservation equation since $\nabla _{\mu }$ is
a covariant derivative. The usual conservation equation with ordinary
derivative reads:

\[
\partial _{\mu }\,\tau ^{\mu \nu }=0 
\]
where $\tau _{\mu \nu }$, which is not a tensor, is defined as \cite{Misner}:

\begin{equation}
\tau ^{\mu \nu }=T_{eff}^{\mu \nu }=\ T^{\mu \nu }+t^{\mu \nu }  \label{tau}
\end{equation}
where $t_{\mu \nu }$ reads: 
\[
\ t_{\mu \nu }=\frac{1}{16\pi }\left[ {\cal L}\,g_{\mu \nu }-\frac{\partial 
{\cal L}}{\partial g_{\mu \nu },\lambda }\ g_{\mu \nu },\lambda \right] 
\]
where ${\cal L}$ is the system's Lagrangian of the following equation: 
\[
T^{\mu \nu }(x)=\frac{\delta S}{\delta g_{\mu \nu }(x)}=\frac{\delta }{%
\delta g_{\mu \nu }(x)}\int {\cal L}\sqrt{-g}dx^{4} 
\]
$t_{\mu \nu }$ is also an homogeneous and quadratic function of the
connection $\Gamma _{\nu \mu }^{\lambda }$ \cite{Landau}, precisely: 
\[
t^{\mu \nu }=\frac{1}{16\pi k}\left[ \left( 2\Gamma _{\alpha \beta }^{\chi
}\Gamma _{\chi \delta }^{\delta }-\Gamma _{\alpha \delta }^{\chi }\Gamma
_{\beta \chi }^{\delta }-\Gamma _{\alpha \chi }^{\chi }\Gamma _{\beta \delta
}^{\delta }\right) (g^{\mu \alpha }g^{\nu \beta }-g^{\mu \nu }g^{\alpha
\beta })+\right. 
\]
\[
g^{\mu \alpha }g^{\beta \chi }\left( \Gamma _{\alpha \delta }^{\nu }\Gamma
_{\beta \chi }^{\delta }+\Gamma _{\beta \chi }^{\nu }\Gamma _{\alpha \delta
}^{\delta }-\Gamma _{\chi \delta }^{\nu }\Gamma _{\alpha \beta }^{\delta
}-\Gamma _{\alpha \beta }^{\nu }\Gamma _{\chi \delta }^{\delta }\right)
+g^{\nu \alpha }g^{\beta \chi }\left( \Gamma _{\alpha \delta }^{\mu }\Gamma
_{\beta \chi }^{\delta }+\Gamma _{\beta \chi }^{\mu }\Gamma _{\alpha \delta
}^{\delta }-\Gamma _{\chi \delta }^{\mu }\Gamma _{\alpha \beta }^{\delta
}-\Gamma _{\alpha \beta }^{\mu }\Gamma _{\chi \delta }^{\delta }\right) + 
\]
\begin{equation}
\left. g^{\alpha \beta }g^{\chi \delta }\left( \Gamma _{\alpha \chi }^{\mu
}\Gamma _{\beta \delta }^{\nu }-\Gamma _{\alpha \beta }^{\mu }\Gamma _{\chi
\delta }^{\nu }\right) \right]  \label{Landau}
\end{equation}
Now we can consider the coordinates $\tau ^{0\mu }$, which satisfy a usual
conservation equation: 
\[
\partial _{\mu }\tau ^{0\mu }=\partial _{0}\tau ^{00}+\partial _{i}\tau
^{0i}=0 
\]
where $\tau ^{00}$ represents the energy density and the $\tau ^{0i}$ are
the three components of the spatial energy flow (the Poynting vector).

Analogously to the case of the energy-momentum tensor, in any orthonormal
basis $\left\{ V_{(\alpha )}^{\mu }\right\} $, $\tau ^{0\alpha }V_{(\alpha
)}^{\mu }$ is not a vector since it is not transformed as a vector by the
Lorentz transformations. But in contrast to that case, the dominant energy
condition cannot be expressed in terms of $\tau ^{\mu \nu }$. However, this
can be done in the particular case of {\it local inertial frames}. In fact,
in a local inertial frame, $\Gamma _{\nu \mu }^{\lambda }=0$; thus, $t_{\mu
\nu }=0$ (see eq.(\ref{Landau})) and $\tau ^{\mu \nu }=$ $T^{\mu \nu }$ (see
eq.(\ref{tau})). Therefore, in this case the dominant energy condition
implies that:

\[
\tau ^{00}\geq \left| \tau ^{\alpha \beta }\right| \qquad
for\,\,each\,\,\alpha ,\beta 
\]
As a consequence, in the basis $\left\{ W_{(\alpha )}^{\mu }\right\} $ $%
=(W_{(0)}^{\mu },W_{(1)}^{\mu },W_{(2)}^{\mu },W_{(3)}^{\mu })$ of the local
inertial frame, the energy flow, represented by $\tau ^{0\alpha }W_{(\alpha
)}^{\mu }$ and satisfying the usual conservation equation, is non-spacelike.

Although this result cannot be generalized to all the reference frames of
the whole spacetime, it is relevant in local contexts since, in small
regions of the spacetime, the metric tends to the Minkowski form\footnote{%
Near each point $x_{0}$, the metric can be approximated with the metric of
the free inertial frame as: 
\[
ds^{2}=(1-R_{0l0m}x^{l}x^{m})dt^{2}-\left( \frac{4}{3}R_{0ljm}x^{l}x^{m}%
\right) dtdx^{j} 
\]
\[
+\left( \delta _{ij}-\frac{1}{3}R_{iljm}x^{l}x^{m}\right)
dx^{i}dx^{j}+0(|x^{j}|^{3})dx^{\alpha }d^{\beta } 
\]
(see \cite{Misner} eq. (13.73)). Therefore, locally in the inertial frame at 
$x_{0}$, we can be sure that $\tau ^{00}\geq |\tau ^{0i}|$.}. In fact, any
local region of the spacetime can be approximated to the tangent Minkowski
space (with orthonormal basis $\left\{ V_{(\alpha )}^{\mu }\right\} $) at
any point of that region. If $\left\{ W_{(\alpha )}^{\mu }\right\} $ is the
basis of an inertial frame on this flat tangent space, $\tau ^{0\alpha
}W_{(\alpha )}^{\mu }$ represents the non-spacelike {\it local energy flow}
satisfying the usual conservation equation. Moreover, at each point $x$ of
the local region, the local energy flow $\tau ^{0\alpha }(x)W_{(\alpha
)}^{\mu }(x)$ belongs to the same light semicone than the one to which $%
T^{0\alpha }(x)V_{(\alpha )}^{\mu }(x)$ belongs. Therefore, if we have
adopted the usual physical convention in the global level, we can also
meaningfully say that future is the temporal direction of the positive local
energy flow: the local flow emitted at $x$ belongs to the light semicone $%
C_{+}(x)$.

This result is particularly relevant because local inertial frames are the
reference frames in which the non-relativistic local theories of physics are
valid. In turn, ordinary quantum field theory in flat spacetime must be
considered as locally formulated in a local inertial frame. This means that
the local energy flow directed towards the future is the flow of energy as
conceived by this kind of theories, where energy satisfies the usual
conservation law expressed by means of ordinary derivatives. Summing up, $%
\tau ^{0\mu }$ inherits the time-orientation defined at the global level
and, to the extent that it has a local physical meaning, it not only
transfers the global arrow of time to local contexts, but also translates
the global arrow into a usual magnitude of local physical theories.

\subsection{The absolute nature of the arrow of time}

As we have seen, the vector $\tau ^{0\mu }$ {\it is always non-spacelike}
and, therefore, its direction defines the arrow of time. However, we know
that, given the time-reversal invariance of Einstein's field equations, in a
time-orientable spacetime where $t$ is the cosmic time, if $%
h_{ij}(t,x^{1},x^{2},x^{3})$ is a solution, $h_{ij}(-t,x^{1},x^{2},x^{3})$
is also a solution; the first case corresponds to $\tau ^{0\mu }$ and the
second case corresponds to $-\tau ^{0\mu }$. At this point, the ghost of
symmetry threatens again: it seems that we are committed to supplying a
non-conventional criterion for picking out one of both nomologically
admissible solutions, one the temporal mirror image of the other.
Nevertheless, as we will see, the threaten is not so serious as it seems.

To replace $t$ by $-t$ is to apply a symmetry transformation, in particular,
time-reversal. The point is to understand the meaning of such a
transformation. Under the active interpretation, a symmetry transformation
corresponds to a change from one system to another; under the passive
interpretation, a symmetry transformation consists in a change of the point
of view from which the system is described. The traditional position about
symmetries assumes that, in the case of discrete transformations, only the
active interpretation makes sense: an idealized observer can rotate himself
in space in correspondence with the given spatial rotation, but it is
impossible to ''rotate in time'' (see \cite{Earman}, \cite{Sklar}). Of
course, this is true when the idealized observer is immersed in the same
spacetime as the observed system. But when the system is the universe as a
whole, we cannot change our spatial perspective with respect to the
universe: it is so impossible to rotate in space as to rotate in time.
However, this does not mean that the active interpretation is the correct
one: the idea of two identical universes, one translated in space or in time
regarding the other, has no meaning. This shows that both interpretations,
when applied to the universe as a whole, collapse into conceptual nonsense.

In fact, in cosmology symmetry transformations are neither given an active
nor a passive interpretation: two mathematical models for the universe,
defined by $(M,g)$\ and $(M^{\prime },g^{\prime })$, are taken to be
equivalent if they are isometric, that is, if there is a diffeomorphism $%
\theta :M\rightarrow M^{\prime }$\ which carries the metric $g$\ into the
metric $g^{\prime }$\ (see \cite{Hawking-Ellis}). Since symmetry
transformations are isometries, two models related by a symmetry
transformation (in particular, time-reversal) are considered equivalent
descriptions of one of the same universe. Therefore, it is not necessary to
find a non-conventional criterion for selecting one of two nomologically
admissible solutions to the extent that both are descriptions of a single
possible universe.

These considerations point to the absolute character of the arrow of time
embodied in $\tau ^{0\mu }$. This vector (in particular, its direction) is
rigidly linked to the spacetime on which it is defined. To change $\tau
^{0\mu }$\ by $-\tau ^{0\mu }$\ amounts to change the model by its temporal
mirror image; but, as we have shown, this move has no physical meaning since
both models are merely different descriptions of the same universe.

Let us note that we have not used the term ''future'' in the present
argument. If we adopt the usual terminology, we will call the time direction
of the energy flow ''future'': in this case, we can say that the vector $%
\tau ^{0\mu }$ points to the future. Nevertheless, it is worth remembering
that ''past'' and ''future'' are words coming from our everyday language but
they do not belong to physical theories. Then, the choice of saying that $%
\tau ^{0\mu }$ points to future is conventional: we can replace ''future''
by ''past'' and nothing changes. What remains is the absolute and
substantial nature of the arrow of time defined by the unchangeable
direction of $\tau ^{0\mu }$.

\subsection{Breaking the symmetry in time-reversal invariant theories}

As we have seen in subsection 3.A, the Ehrenfest's objected Gibbs' approach
by pointing out that the increase of the entropy towards the future is
always matched by a similar one in the past of the system. It is interesting
to note that this old discussion can be generalized to the case of any kind
of evolution arising from local time-reversal invariant laws. In fact,
time-reversal invariant equations always give rise to what we will call ''%
{\it time-symmetric twins'' }(see \cite{Olimpia}), that is, two mathematical
structures symmetrically related by a time-reversal transformation: each
''twin'', which in some cases represents an irreversible evolution, is the
temporal mirror image of the other ''twin''. For instance, electromagnetism
provides a pair of advanced and retarded solutions, that are usually related
with incoming and outgoing states in scattering situations as described,
e.g., by Lax-Phillips scattering theory \cite{Lax-Phillips}. In irreversible
quantum mechanics, the analytical extension of the energy spectrum of the
quantum system's Hamiltonian into the complex plane leads to poles in the
lower half-plane (usually related with decaying unstable states), and
symmetric poles in the upper half-plane (usually related with growing
unstable states) (see \cite{Cast-Laura}), etc. However, at this level the
twins are only conventionally different: we cannot distinguish between
advanced and retarded solutions or between lower and upper poles without
assuming temporally asymmetric notions, as the asymmetry between past and
future or between preparation and measurement. Then, the challenge consists
in supplying a non-conventional criterion for choosing one of the twins as
the physically relevant: such a criterion must establish a substantial
difference between the two members of the pair.

The arrow of time defined by $\tau ^{0\mu }$ is precisely what provides us
the criterion for establishing the desired substantial difference. In fact, $%
\tau ^{0\mu }$ says that at each point of the spacetime the semicones $%
C_{-}(x)$ receive an incoming flow of energy while the semicones $C_{+}(x)$
emit an outgoing flow of energy. Therefore, in each case the twin that must
be retained as physically relevant is the one describing this kind of energy
flow, from $C_{-}(x)$ to $C_{+}(x)$. For instance, in electromagnetism only
retarded solutions fulfill this condition since they describe waves
propagating into the semicone $C_{+}(x)$. In irreversible quantum mechanics,
only decaying unstable states{\bf \ }with poles in the lower half-plane have
the physical sense of spontaneous evolutions, since they provide an energy
flow contained in $C_{+}(x)$. On the contrary, the poles in the upper
half-plane represent growing unstable states, which are not spontaneous
since they must be generated by pumping energy coming from the semicone $%
C_{-}(x)$, that is, from the past. Summing up, by translating the
geometrical global time-asymmetry in terms of energy flow, $\tau ^{0\mu }$
can be used locally (at each point of the spacetime) for breaking the
symmetry of the set of solutions produced by time-reversal invariant laws
(see \cite{Olimpia} for more cases).

\section{The non time-reversal invariance of quantum field theory: the
second role of the energy-momentum tensor}

\subsection{The non time-reversal invariance of axiomatic QFT}

The Postulate A.3 of the axiomatic quantum field theory (see \cite{Haag},
p.56, eq.II.1.15) states that the spectrum of the energy-momentum operator $%
P^{\mu }$ is confined to a future light semicone, that is, its eigenvalues $%
p^{\mu }$ satisfy: 
\[
p^{2}\geq 0\qquad \qquad p^{0}\geq 0 
\]
This postulate says that, when we measure the observable $P^{\mu }$, we
obtain a {\it non-spacelike} {\it classical} $p^{\mu }$ contained in a
future semicone, that is, a semicone belonging to $C_{+}$.

It is clear that condition $p^{0}\geq 0$ selects one of the elements of the
time-symmetric twins, $p^{0}\geq 0$ and $p^{0}\leq 0$ that would arise from
the theory: by means of Postulate A.3, QFT becomes a non time-reversal
invariant theory. In turn, since QFT, being both quantum and relativistic,
can be considered one of the most basic theories of physics, the choice
introduced by condition $p^{0}\geq 0$ is transferred to the rest of physical
theories. But such a choice is established from the very beginning, as a
postulate of the theory. The challenge is, then, to {\it justify} Postulate
A.3 by means of independent theoretical arguments.

As it is well known, the components of $T^{\mu \nu }(x)$ can be interpreted
as follows: 
\begin{eqnarray*}
&&\ \ \ \ \ \ \ \ T^{00}(x)\text{ is the matter-energy density} \\
&&\ \ \ \ \ \ \ \ T^{0i}(x)\text{ is the matter-energy flow} \\
&&\ \ \ \ \ \ \ \ T^{i0}(x)\text{ is the linear momentum density} \\
&&\ \ \ \ \ \ \ \ T^{ij}(x)\text{ is the linear momentum flow}
\end{eqnarray*}
where ($i,j=1,2,3$). Since $T^{\mu \nu }$ is a symmetric tensor, $T^{\mu \nu
}(x)=T^{\nu \mu }(x)$ and, therefore, $T^{0i}(x)=T^{i0}(x)$; in other words,
the matter-energy flow is equal to the linear momentum density. This means
that. if the matter energy flow $T^{0\alpha }$ can be used to define the
arrow of time under the dominant energy condition, this is also the case for
the linear momentum density $T^{\alpha 0}$. On the other hand, we have seen
in Subsection 5.B that the local matter-energy flow $\tau ^{\mu \nu }$ can
be conceived as a conservative version of $T^{\mu \nu }$ in the orthonormal
coordinates of a local inertial frame; in this case, the dominant energy
condition has the consequence that $\tau ^{0\mu }$ is non-spacelike. Now we
know that exactly the same conclusion can be drawn for the local linear
momentum density $\tau ^{\mu 0}$. But it is precisely the local linear
momentum density $\tau ^{\mu 0}$ the local magnitude corresponding to the
classical {\bf \ }$p^{\mu }$ of QFT; thus, at each point $x$, $p^{\mu }\sim
\tau ^{0\mu }\in C_{+}(x)$.

In conclusion, the fact that {\bf \ }$p^{\mu }$ at each point $x$ of the
local context and, therefore, for every classical particle, must be
contained in the future light semicone $C_{+}(x)$ turns out to be a
consequence of the global time-asymmetry of the spacetime when the dominant
energy condition holds everywhere. In other words, Postulate A.3 can be
justified on cosmological grounds instead of being imposed as a departing
point of the axiomatic version of QFT.

\subsection{The non time-reversal invariance of ordinary QFT}

In this section we will analyze how time-reversal invariance is introduced
in the ordinary version of QFT. In order to do this let us remember again
that, in the tangent plane at each point $x$ of our manifold, we can define
an orthonormal tetrad $V_{(\alpha )}^{\mu }=\{V_{(0)}^{\mu },V_{(i)}^{\mu }\}
$, where $V_{(0)}^{\mu }$ is a timelike vector and the $V_{(i)}^{\mu }$ are
spacelike vectors.

\begin{quote}
{\bf Definitions 13:} The complete Lorentz group $L$ has four components. $%
L_{0}$ is the identity component, also known as the {\it proper} Lorentz
group. The space inversion ${\cal P}$, defined as $(V_{(0)}^{\mu
}\rightarrow V_{(0)}^{\mu },V_{(i)}^{\mu }\rightarrow -V_{(i)}^{\mu })$, is
the typical element of the component $L_{1}$. The combination of $L_{0}$ and 
$L_{1}$ is known as the {\it orthochronous} Lorentz group. Analogously, the
time inversion ${\cal T}$ , defined as $(V_{(0)}^{\mu }\rightarrow
-V_{(0)}^{\mu },V_{(i)}^{\mu }\rightarrow V_{(i)}^{\mu })$, is the typical
elements of the component $L_{2}$. The combination of $L_{0}$ and $L_{2}$ is
known as the {\it orthospatial} Lorentz group. Finally, the total inversion $%
(V_{(0)}^{\mu }\rightarrow -V_{(0)}^{\mu },V_{(i)}^{\mu }\rightarrow
-V_{(i)}^{\mu })$ is the typical element of the component $L_{3}$. The
combination of $L_{0}$ and $L_{3}$ is known as the {\it unimodular} Lorentz
group.
\end{quote}

The classification of one-particle states according to their transformation
under the Lorentz group leads to six classes of four-momenta (see \cite
{Weinberg}). Of these classes, it is considered that only three have
physical meaning: these are precisely the cases which agree with Postulate
A.3 of the axiomatic version of QFT. In other words, the symmetry group of
QFT is the orthochronous group, where ${\cal P}$ but not ${\cal T}$ are
included. This is another way of expressing the non time-reversal invariance
of the QFT. In this case, the non time-reversal invariance is introduced not
by means of a postulate but on the basis of empirical arguments that make
physically meaningless certain classes of four-momenta. However, to the
extent that special relativity and standard quantum mechanics are
time-reversal invariant theories, they give no theoretically grounded
justification for such a breaking of time-reversal invariance. Nevertheless,
as we have seen in the previous subsection, this justification can be given
on cosmological grounds.

Let us make the point in different terms. The quantum field correlates of $%
{\cal P}$ and ${\cal T}$ , ${\bf P}$ and ${\bf T}$, are defined as: 
\[
{\bf P}iP^{\nu }{\bf P}^{-1}=i{\cal P}_{\mu }^{\nu }P^{\mu }\qquad \qquad 
{\bf T}iP^{\nu }{\bf T}^{-1}=i{\cal T}_{\mu }^{\nu }P^{\mu } 
\]
where ${\bf P}$ is a linear and unitary operator and ${\bf T}$ is an
antilinear and antiunitary operator. In fact, if ${\bf T}$ were linear and
unitary, we could simply cancel the $i$'s and, then, ${\bf T}P^{\nu }{\bf T}%
^{-1}=-P^{\mu }$: the action of the operator ${\bf T}$ on the operator would
invert the sign of $P^{\mu }$, with the consequence that the spectrum of the
inverted energy-momentum operator would be contained in a past light
semicone. In particular, for $\nu =0$, $P^{\mu }=H$, where $H$ is the energy
operator; then, if ${\bf T}$ were linear and unitary, ${\bf T}H{\bf T}%
^{-1}=-H$ with the consequence that, for any state of energy $E$ there would
be another state of energy $-E$. The antilinearity and antiunitarity of $%
{\bf T}$ avoid these ''anomalous'' situations in agreement with the
conditions imposed by Postulate A.3 and, at the same time, make QFT non
time-reversal invariant. Once again, there are good empirical reasons for
making ${\bf T}$ antilinear and antiunitary, but not theoretical
justification for such a move.

Summing up, in ordinary QFT it is always necessary to take a decision about
the time direction of the spectrum of the energy-momentum operator $P^{\mu }$%
. The point that we want to stress here is that, as in the case of Postulate
A.3 of the axiomatic version of QFT, the decision can be justified on
cosmological grounds, as a consequence of the global time-asymmetry of the
universe and the dominant energy condition.

Finally, it is worth reflecting on the role of weak interactions in the
problem of the arrow of time. The CPT theorem states that ${\bf CPT}$ is the
only combination of charge conjugation ${\bf C}$, parity reflection ${\bf P}$
and time-reversal ${\bf T}$ which is a symmetry of QFT. In fact, it is well
known that weak interactions break the ${\bf T}$ of the CPT theorem.
According to a common opinion, it is precisely this empirical fact the clue
for the solution of the problem of the arrow of time: since the ${\bf T}$
symmetry is violated by weak interactions, they introduce a non-conventional
distinction between the two directions of time (see \cite{Visser}). The
question is: is the breaking of ${\bf T}$ what distinguishes both directions
of time in QFT? As we have seen, the operator ${\bf T}$ was designed
precisely to avoid that certain tetra-magnitudes, such as the linear
momentum $p^{\mu }$, have the ''anomalous'' feature of being contained in a
past light semicone: the action of the operator ${\bf T}$ onto the
energy-momentum operator $P^{\mu }$ conserves the time direction of $P^{\mu }
$ and, therefore, of its eigenvalues. It is this fundamental fact what makes
QFT non time-reversal invariant, and not the incidental violation of ${\bf T}
$ by weak interactions\footnote{%
Of course, this leaves open a different problem: to explain why, among all
the elementary interactions, only weak interactions break ${\bf T}$.}. This
non time-reversal invariance of QFT, based on the peculiar features of ${\bf %
T}$, distinguishes by itself between the two directions of time, with no
need of weak interactions. In other words, even if weak interactions did not
exist, QFT would be a non time-reversal invariant theory which would define
the arrow of time. The real problem is, then, to justify the non
time-reversal invariance of a theory which is presented as a synthesis of
two time-reversal invariant theories such as special relativity and quantum
mechanics. But this problem is completely independent of the existence of
weak interactions and the breaking of ${\bf T}$ introduced by them. Summing
up, weak interactions do not play a role as relevant in the problem of the
arrow of time as it is usually supposed.

\section{ Conclusion}

In this paper we have defined a physical object $\tau ^{0\mu }$ $\sim p^{\mu
}$ that can play the role of the arrow of time. Then, the mysterious and
phantomlike arrow of Eddington is at last materialized. In particular, we
have shown the double role played by the energy-momentum tensor in the
context of our approach to the problem of the arrow of time. When the
matter-energy flow is considered, the energy-momentum tensor translates the
geometrical time-asymmetry of the universe in terms of energy flow. When the
linear momentum density is considered, the energy-momentum tensor provides
the means of justifying the time-asymmetry postulate of axiomatic quantum
filed theory.


\begin{references}
\bibitem{Casta Varios}  M. Castagnino, F. Gaioli and E. Gunzig, {\it Found.
Cosmic Phys.}, {\bf 16, }221, 1996.

M. Castagnino and R. Laura, {\it Phys. Rev. A}, {\bf 56}, 108, 1997.

M. Castagnino, ''The global nature of time asymmetry and the
Bohm-Reichenbach diagram''{\it , }in A. Bohm, H. Doebner and P. Kielarnowski
(eds), {\it Irreversibility and Causality (Proc. G. 21 Goslar 1996)},
Springer-Verlag, Berlin, 1998.

M. Castagnino and E. Gunzig, {\it Int. Journ. Theo. Phys}. {\bf 38,} 47,
1999.

M. Castagnino and C. Laciana, {\it Class. Quant. Grav.}, {\bf 19,} 2657,
2002.

M. Castagnino, J. Gueron and A. Ordo\~{n}ez, {\it J. Math. Phys}., {\bf 43},
705, 2002.

M. Castagnino, G. Catren and R. Ferraro, {\it Class. Quant. Grav}., {\bf 19}%
, 4729, 2002.

M. Castagnino, L. Lara and O. Lombardi, ''The direction of time: from global
arrow to local arrow'',{\it \ Int. Jour. Theo. Phys}., at press, 2003.

\bibitem{Olimpia}  M. Castagnino, L. Lara and O. Lombardi, {\it Class.
Quant. Grav.}, {\bf 20}, 369, 2003.

\bibitem{Foundations}  M. Castagnino, O. Lombardi and L. Lara, {\it Found.
Phys}., {\bf 33}, 857, 2003.

\bibitem{Irrev}  C. G. Sudarshan, C. B. Chiu and V. Gorini, {\it Phys. Rev. D%
}, {\bf 18, }2914, 1978.

A. Bohm,{\it \ Quantum Mechanics, Foundations and Applications},
Springer-Verlag, Berlin, 1979.

A. Bohm and M. Gadella, {\it Dirac Kets, Gamow Vectors, and Gel'fand triplets%
}, Springer- Verlag, Berlin, 1989.

M. Castagnino and E. Gunzig, {\it Int. Jour. Theo. Phys}., {\bf 36}, 2545,
1997.

\bibitem{Penrose}  R. Penrose, ''Singularities and time asymmetry'', in S.
Hawking and W. Israel (eds.), {\it General Relativity, an Einstein Centenary
Survey}, Cambridge University Press, Cambridge, 1979.

\bibitem{Sachs}  R. G. Sachs, {\it The Physics of Time-Reversal}, University
of Chicago Press, Chicago, 1987.

\bibitem{Price}  H. Price, {\it Time's Arrow and the Archimedes' Point, }%
Oxford University Press, Oxford, 1996.

\bibitem{Ehrenfest}  P. Ehrenfest and T. Ehrenfest, {\it The Conceptual
Foundations of the Statistical Approach in Mechanics}, Cornell University
Press, Ithaca, 1959 (original 1912).

\bibitem{Boltzmann}  L. Boltzmann, {\it Annalen der Physik}., {\bf 60, }392,
1897.

\bibitem{Feynman}  R. P. Feynman, R. B. Leighton and M. Sands, {\it The
Feynman Lectures on Physics, Vol. 1, }Addison-Wesley{\it , }New York, 1964.

\bibitem{Davies}  P. C. Davies, ''Stirring up trouble'', in J. J. Halliwell,
J. Perez-Mercader and W. H. Zurek (eds.), {\it Physical Origins of Time
Asymmetry}, Cambridge University Press, Cambridge, 1994.

\bibitem{Earman}  J. Earman, {\it Phil. Scien}., {\bf 41, }15, 1974.

\bibitem{Matthews}  G. Mattews, {\it Phil. Scien}., {\bf 46, }82, 1979.

\bibitem{Schul}  S. Schulman, {\it Phys. Rev. Lett.}, {\bf 83,} 5419, 1999.

\bibitem{Hawking-Ellis}  S. Hawking and J. Ellis, {\it The Large Scale
Structure of Space-Time, }Cambridge University Press, Cambridge, 1973.

\bibitem{Schutz}  B. F. Schutz, {\it Geometrical Methods of Mathematical
Physics}, Cambridge University Press, Cambridge, 1980.

\bibitem{Misner}  C. W. Misner, K. S. Thorne and J. A Wheeler, {\it %
Gravitation}, Freeman and Co., San Francisco, 1973.

\bibitem{WMAP}  C. L. Bennet, M. Halpern, G. Hinshaw, N. Jarosik, A. Kogut,
M. Limon, S. S. Meyer, L. Page, D. N. Spergel, G. S. Tucker, E. Wollack, E.
L. Wright, C. Barnes, M. R. Greason, R. S. Hill, E. Komatsu, M. R. Nolta, N.
Odegard, H. V. Peirs, L. Verde and J. L. Weiland, ''First year Wilkinson
Microwave Anisotropy Probe (WMAP)'', {\it arXiv:astro-ph/0302207.}

A. Kogut, D. N. Spergel, C. Barnes, C. L. Bennet, M. Halpern, G. Hinshaw, N.
Jarosik. M. Limon, S. S. Meyer, L. Page, G. Tucker, E. Wollack and E. L.
Wright, ''Wilkinson Microwave Anisotropy Probe (WMAP). First year '', {\it %
arXiv:astro-ph/0302213.}

\bibitem{SUIe}  S. Perlmutter and B.\ P. Schmidt, ''Measuring cosmology with
supernovae'', {\it arXiv:astro-ph/0303428.}

\bibitem{Cast-Giac-Lara}  M. Castagnino, H. Giacomini and L. Lara, {\it %
Phys. Rev. D}, {\bf 61}, 107302, 2000; {\bf 63}, 044003, 2001.

\bibitem{Savitt}  S. F. Savitt, {\it Brit. Jour. Phil. Scie.}, {\bf 47},
347, 1996.

\bibitem{Visser}  M. Visser, {\it Lorentzian Wormholes}, Springer-Verlag,
Berlin, 1996.

\bibitem{Landau}  L. Landau and E. Lifchitz, {\it Th\'{e}orie des Champs},
Editions Mir, Moscow, 1970.

\bibitem{Sklar}  L. Sklar, {\it Space, Time and Spacetime}, University of
Califormia Press, Berkeley, 1974

\bibitem{Lax-Phillips}  P. D. Lax and R. S. Phillips,{\it \ Scattering Theory%
}, Academic Press, New York, 1979.

\bibitem{Cast-Laura}  M. Castagnino and R. Laura, {\it Phys. Rev. A}, {\bf 56%
}, 108, 1997.

\bibitem{Haag}  R. Haag, {\it Local Quantum Physics. Fields, Particles,
Algebras}, Springer, Berlin, 1996.

\bibitem{Weinberg}  S. Weinberg, {\it The quantum theory of fields, }%
Cambridge Univ. Press, Cambridge, 1995.
\end{references}
\end{document}